\documentclass[a4paper]{jpconf}
\usepackage{graphicx}
\begin {document}

\title{Stable and habitable systems with two giant planets}
\author{Vera Dobos$^1$, Imre Nagy$^{2,1}$ and Judit Orgov\'anyi$^1$}
\address{$^1$ E\"otv\"os University, P\'azm\'any P\'eter s\'et\'any 1/A, 1117 Budapest, Hungary}
\address{$^2$ Physical Geodesy and Geodesical Research Group of the HAS, Technical University, M\H{u}egyetem rakpart 3, 1111 Budapest, Hungary}
\ead{vera.zs.dobos@gmail.com}

\begin{abstract}
We have studied planetary systems which are similar to the Solar System and built up from three inner rocky planets (Venus, Earth, Mars) and two outer gas giants. The stability of the orbits of the inner planets is discussed in the cases of different masses of the gas planets. To demonstrate the results stability maps were made and it was found that Jupiter could be four times and Saturn could be three times more massive while the orbits of the inner planets stay stable. Similar calculations were made by changing the mass of the Sun. In this case the position of the rocky planets and the extension of the liquid water habitable and the UV habitable zones were studied for different masses of the Sun. It was found that the orbits of the planets were stable for values greater than 0.33 $M_\odot$ where $M_\odot$ is the mass of the Sun and at lower masses of the Sun (at about 0.8 $M_\odot$) only Venus, but for higher mass values (at about 1.2 $M_\odot$) Earth and also Mars are located in both habitable zones.
\end{abstract}

\section{Introduction}
New planetary systems containing one or more gas giants are discovered day by day. Present instruments and methods are, however, not capable of detecting rocky planets, hence these discoveries are considered to be rare events. As instruments are getting more and more developed the number of known terrestrial-mass planets is expected to grow significantly (Csizmadia, 2009). 

Search for extraterrestrial life is also one of the main goals of ESA and NASA. New interferometers of these space agencies are to be launched in the next decade with the aim of discovering and characterizing new Earth-like planets. Therefore it is important to investigate the conditions within which the orbits of these planets are stable in the given system.

What would be the outcome in terms of the stability of the Solar System, if the masses of Jupiter and Saturn had been formed in a different way from the cloud envelope of the protostellar Sun? Is the Solar System unique? In this work we discuss different cases similar to the Solar System to investigate the conditions of habitability on rocky planets.

Our research program consists of two main parts. The first part focuses on calculations related to celestial mechanics that is the stability of the rocky planets is investigated as a function of the masses of the gas giants and as a function of the mass of the Sun. This latter case is closely related to the second part of our work which is the calculation of the extension of the two kinds of habitable zones: the conventional liquid water habitable zone and the ultraviolet habitable zone.

\section{Methods}
\subsection{Stability calculations}
The studied planetary systems are built up from three inner rocky planets: Venus, Earth, Mars and two outer gas giants: $P_1$ and $P_2$, where $P_1$ is closer while $P_2$ is located farther from the Sun. $M_1$ and $M_2$ are the masses of $P_1$ and $P_2$, respectively. By changing the masses of the gas giants from 0 to 10 $M_J$ where $M_J$ is the mass of Jupiter the integration was made for $5000 \: P_S$ ($P_S$: period of Saturn) which is about 150000 years. Initial orbital elements were obtained for J.D. = 2415020.0 epoch where $P_1$ and $P_2$ were chosen for the parameters of Jupiter and Saturn, respectively. Stability of the inner planets is described by the maximum of eccentricities and close encounters. For the latter we investigated several parameters but finally chose the difference between the distance of the outer planet's pericentre and the distance of the inner planet's apocentre.

In the other case the planets' masses were constants and the mass of the Sun was altered between 0.1 $M_\odot$ and 10 $M_\odot$ where $M_\odot$ is the mass of the Sun. The integration time was the same as in the previous case that is 5000 $P_S$. The stability was characterized by the pericentre and apocentre of the planets and both habitable zones were calculated for a few masses. For the latter the radius of the star was calculated from its mass using the formulae constructed by Zaninetti (2008).

\subsection{Habitable zones}
The liquid water habitable zone (LW HZ) is a region around a star in which an Earth-like planet could support liquid water on its surface (Kasting et al., 1993). In the equation that describes the boundaries of the LW HZ the Solar system is used as a reference for the calculations of other planetary systems.

An idea from Buccino et al. (2006) was borrowed that there must be both an inner and an outer boundary for sufficient and necessary UV radiation level respectively (UV HZ). It is well known that this radiation induces DNA\footnote{The deoxyribonucleic acid (DNA) contains the genetic instructions used in the development and functioning of all known living organisms.} damage (especially UVB and UVC), inhibits photosynthesis (UVA) and causes lesion in a wide variety of proteins and lipids (Cockell, 1998). The action spectrum $B(\lambda)$ is commonly used as a measure of damage, quantifying the injurious effects of the UV radiation on the biological processes in function of the wavelength (Coohill, 1991; Horneck, 1995; Cockell, 1998). The action spectrum is determined by exponential functions given by Modos et al. (1999) who measured the spectrum by different dosimeters. The results of uracil dosimeter were chosen for the calculations as uracil is a compound that can be found within the RNA\footnote{The ribonucleic acid (RNA) conveys genetic information and catalyzes important biochemical reactions.}.

The biological effective spectrum for each value of wavelength can be obtained by multiplying the action spectrum with the incident radiation. To get a tractable expression for the calculation of the inner boundary, the Earth was used again as a reference. This way the following inequality was obtained:

\begin {equation}
\frac {d_{inner}^2} {d_\oplus^2}  \ge  \frac { \int\limits_{200nm}^{400nm} { B(\lambda) \cdot {R_*^2} \cdot E_*(\lambda) \: d\lambda } } { x \cdot \int\limits_{200nm}^{400nm} { B(\lambda) \cdot {R_\odot^2} \cdot E_\odot(\lambda) \: d\lambda } } ,
\end {equation}
\\
where $d_{inner}$ and $d_\oplus$ are the distances of the inner boundary and of the Earth, $R$ is the radius and $E(\lambda)$ is the emitted energy of the star and of the Sun at $\lambda$ wavelength and $x$ is the multiplicator of the terrestrial UV radiation that still can be tolerated by DNA, proteins and photosynthetic process. In other words this means that these important compounds can bear even $x$ times of the UV radiation on Earth.

On the other hand the UV radiation is an important energy source needed for chemical synthesis of complex molecules. An expression similar to (1) can be formulated to describe the outer boundary of the UV habitable zone:

\begin {equation}
\frac {d_{outer}^2} {d_\oplus^2}  \le  \frac { \int\limits_{200nm}^{400nm} { {R_*^2} \cdot E_*(\lambda) \: d\lambda } } { y \cdot \int\limits_{200nm}^{400nm} { {R_\odot^2} \cdot E_\odot(\lambda) \: d\lambda } } ,
\end {equation}
\\
where $d_{outer}$ is the distance of the outer boundary and $y$ is the multiplicator of the terrestrial UV radiation which is absolutely necessary for chemical reactions. Obviously $y$ should be set to be less than 1.

The borders of the UV HZ can also be calculated for the ancient Sun since its radiation has increased by about 30\% in the last $4.7 \cdot 10^9$ years (Gough, 1981). In this work the present radiation level was used for the calculations.

\section{Results and discussion}
\subsection{Stability calculations}
\begin{figure}[h]
\centering

\begin{minipage}{18.5pc}
\includegraphics[width=18.5pc]{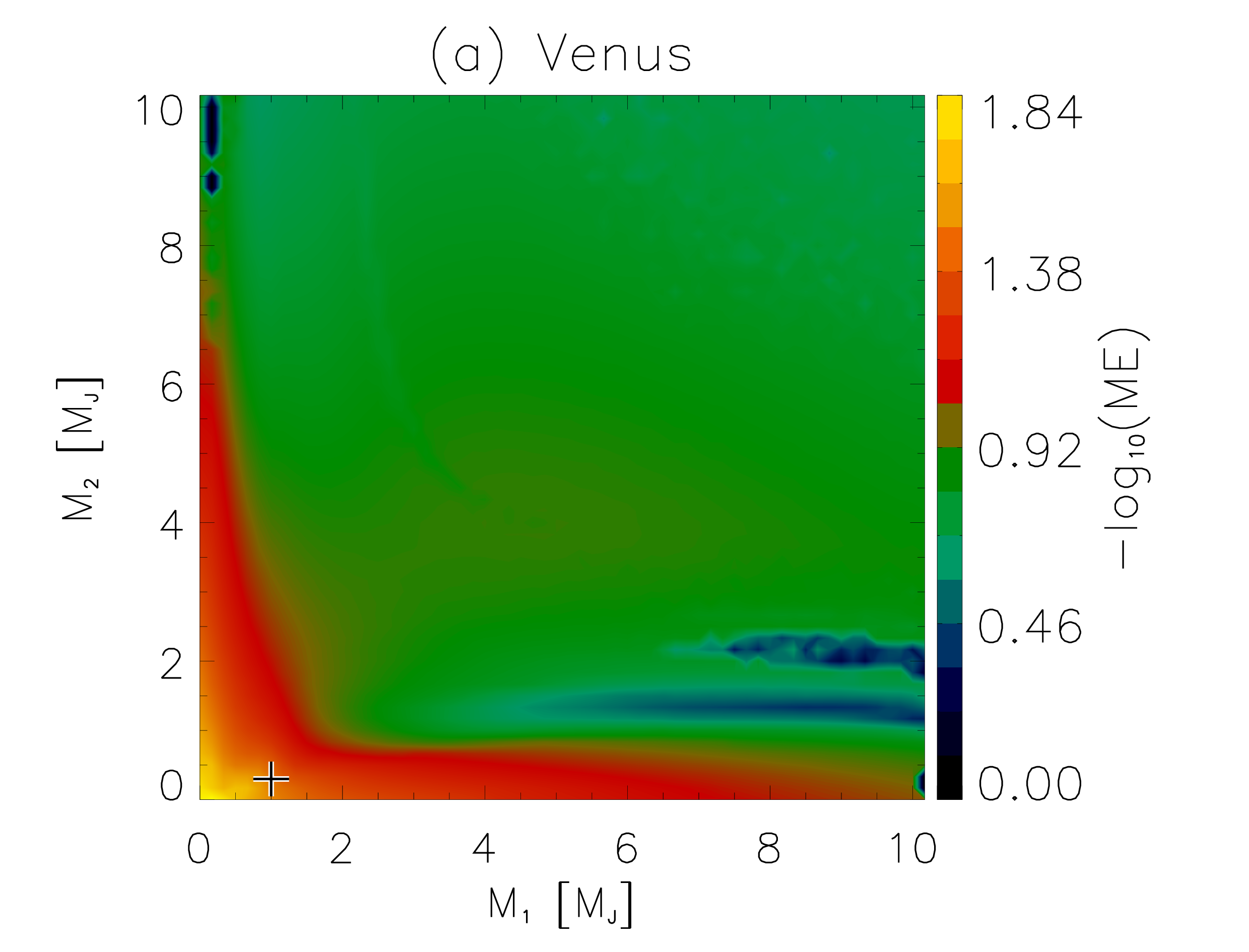}
\end{minipage}
\begin{minipage}{18.5pc}
\includegraphics[width=18.5pc]{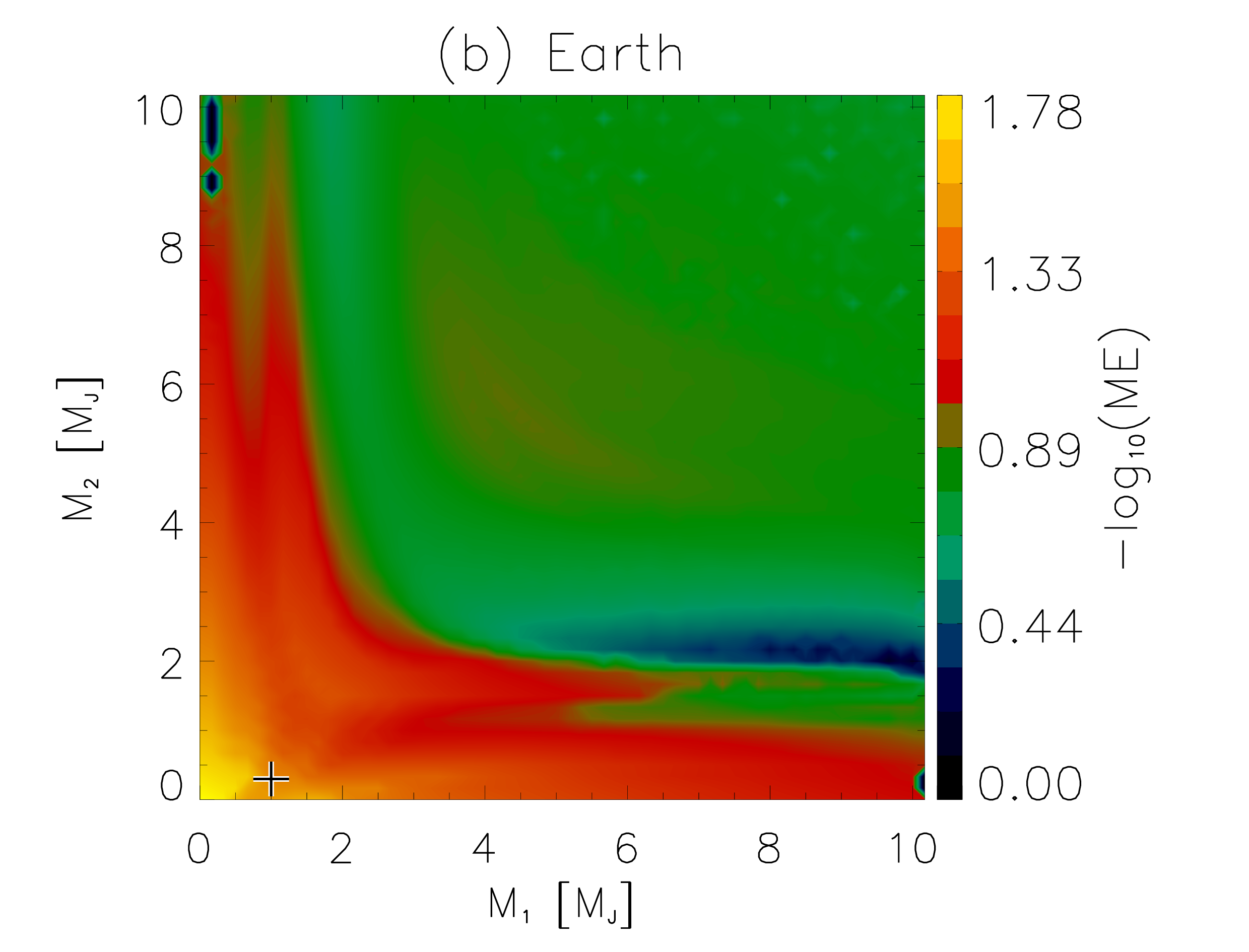}
\end{minipage}
\begin{minipage}{18.5pc}
\includegraphics[width=18.5pc]{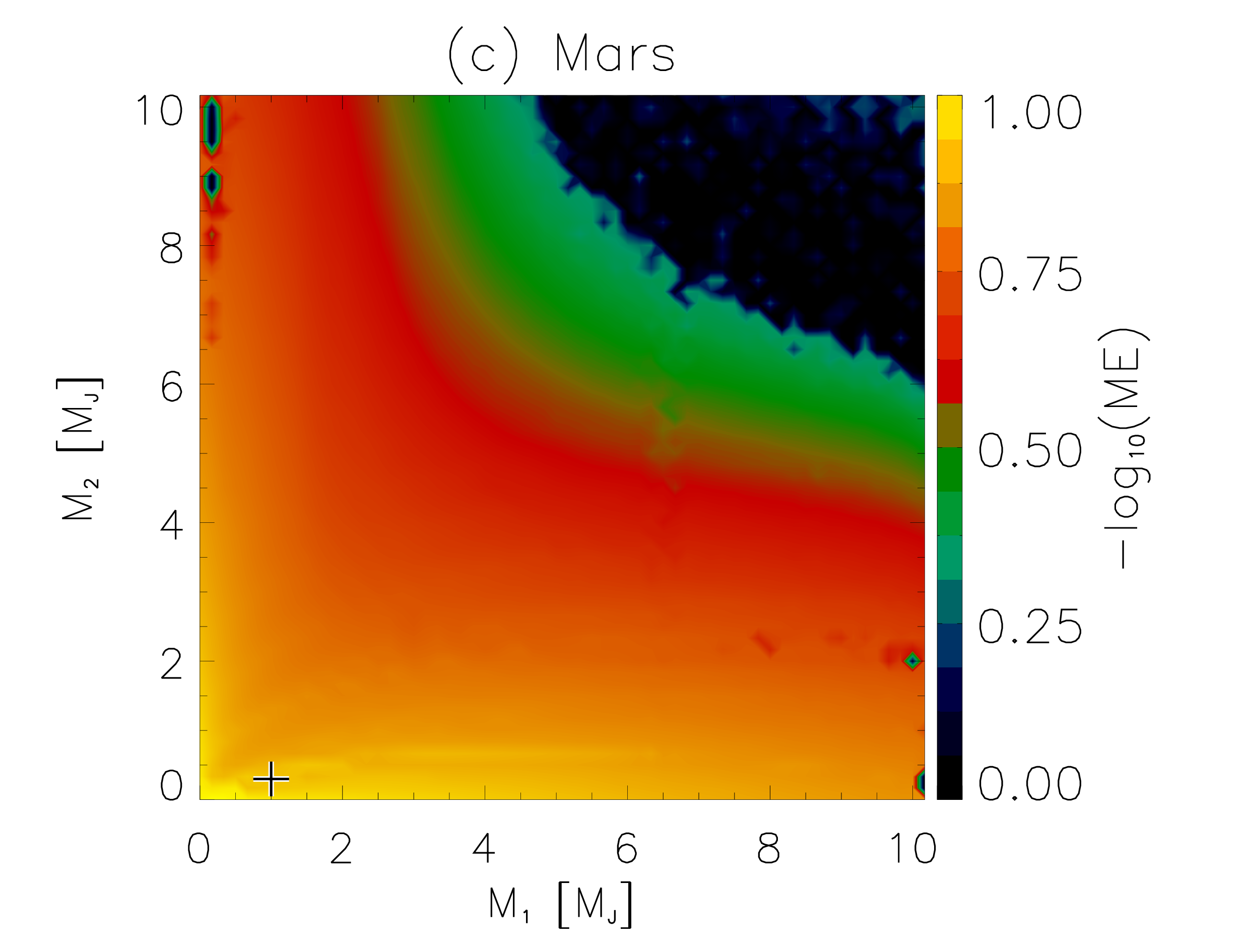}
\end{minipage}
\begin{minipage}{18.5pc}
\includegraphics[width=18.5pc]{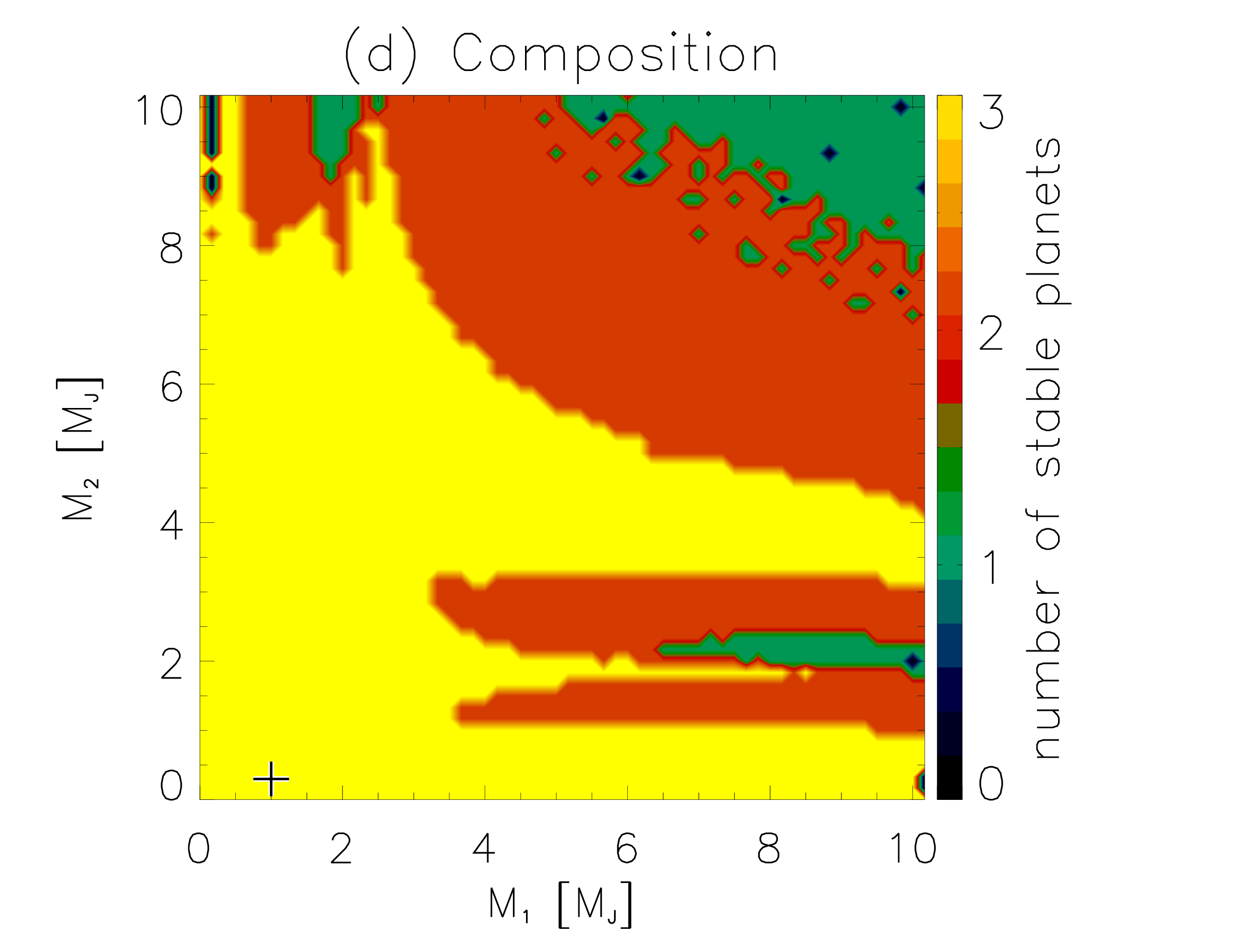}
\end{minipage}
\caption{\label{ex}($a$), ($b$) and ($c$): Stability maps for the three inner planets in function of the masses of $M_1$ (axis $x$) and $M_2$ (axis $y$). ME: maximum value of the eccentricity. Yellow and black colours indicate the most stable and the most unstable regions, respectively. The '+' mark indicates our Solar System. ($d$): Composition of stability maps seen in ($a$) -- ($c$). Colours indicate the number of stable planets: yellow - three, red - two, green - one, black - zero.}

\end{figure}

Stability is investigated by calculating the extrema of the orbital elements (semimajor axis, eccentricity, pericentre and apocentre distance). Using these data stability maps in function of the giant planets' masses ($M_1$ and $M_2$) were plotted (see Fig. \ref{ex} ($a$), ($b$) and ($c$)). Axis $x$ corresponds to the mass $M_1$ while axis $y$ to the mass $M_2$ (both in units of $M_J$). Yellow coloured regions contain those pairs of masses for which the orbits of the rocky planets are the most stable while black colour indicates the most unstable cases. One can see the '+' mark at coordinates (1, 1/3) which corresponds to one Jupiter and one Saturn mass, so this is the case of our Solar System. It can be observed that the figures are almost symmetrical to the $45^\circ$ diagonal line. This is not surprising regarding that the masses of the two giant planets are expressed in the same unit. A slight asymmetry can also be noticed, that is caused by the different distances of the giant planets. For all inner planets the maximum eccentricity is extremely high (black colour) for low $M_1$ mass when $M_2$ is about 9 -- 10 $M_J$. The reason is that in these cases $M_1$ gets closer to the Sun, at about 3 $AU$, thus the inner planets migrate outward. Other conspicuous characteristics are the horizontal lines at 1 -- 2.5 $M_J$ of $M_2$ at high masses of $M_1$. The explanation is that the maximum semimajor axis of Venus gets higher and thus the planet gets closer to Earth. Even in some cases the orbit of Venus partially overlaps the orbit of Mars. If we take a look at the stability map of Mars a big black area is clearly seen at the top right corner. This unstable region is formed by the influence of the two massive giant planets. Here the maximum eccentricity of Mars almost reaches the value of 1.

The planet is considered unstable if the difference between its maximum and original eccentricity is greater than 0.2. Fig. \ref{ex} ($d$) is a composition of the previous stability maps. It shows the number of stable inner planets in function of the masses of the giant planets. Yellow colour means that all three planets are considered stable according to the above criterion while black indicates unstable regions (when all three planets are unstable). The yellow region has a very large extension which means that the orbits of the inner planets are stable for a wide variety of the pairs of the gas giants.

\begin{figure}[h]
\centering

\begin{minipage}{18.5pc}
\includegraphics[width=18.5pc]{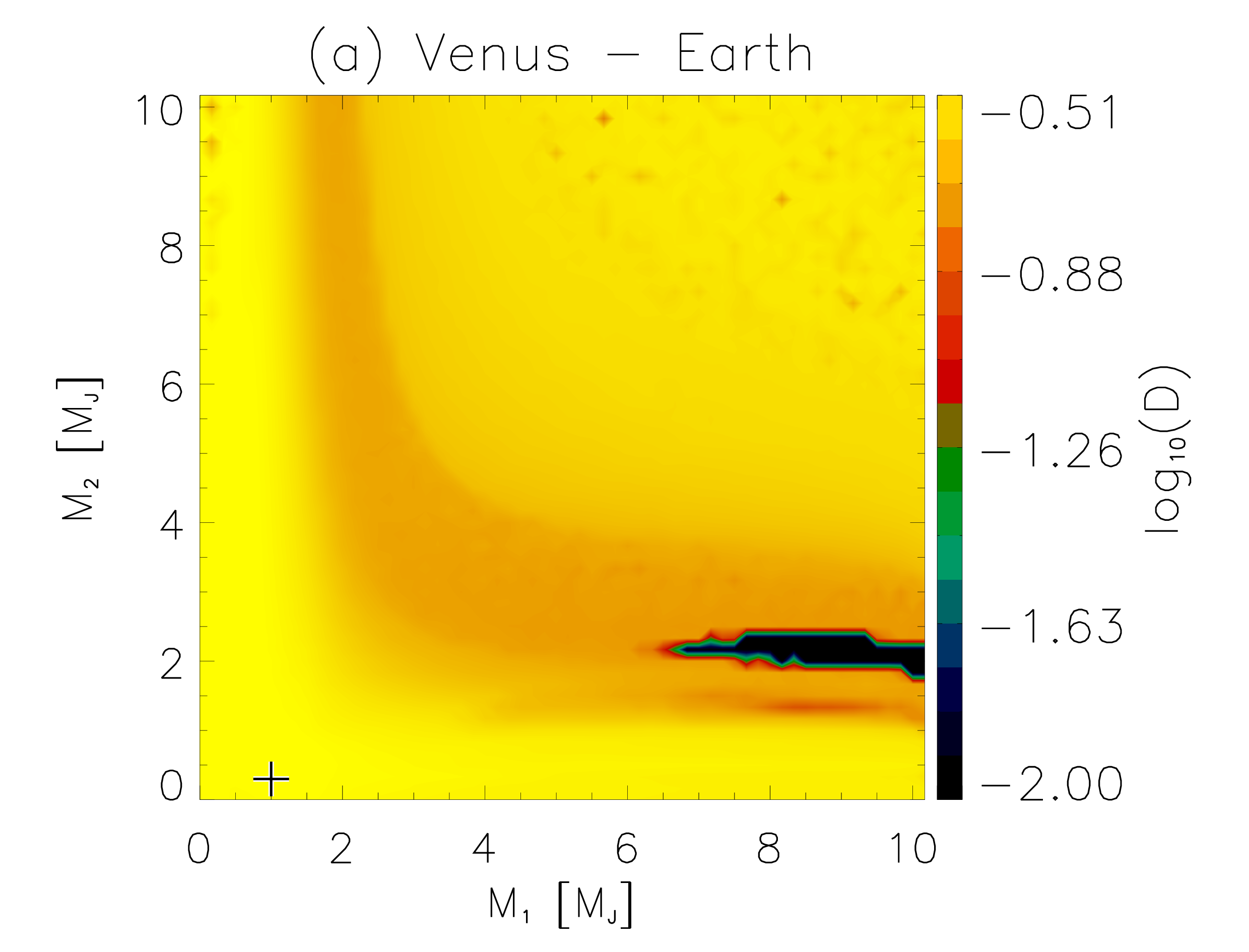}
\end{minipage}
\begin{minipage}{18.5pc}
\includegraphics[width=18.5pc]{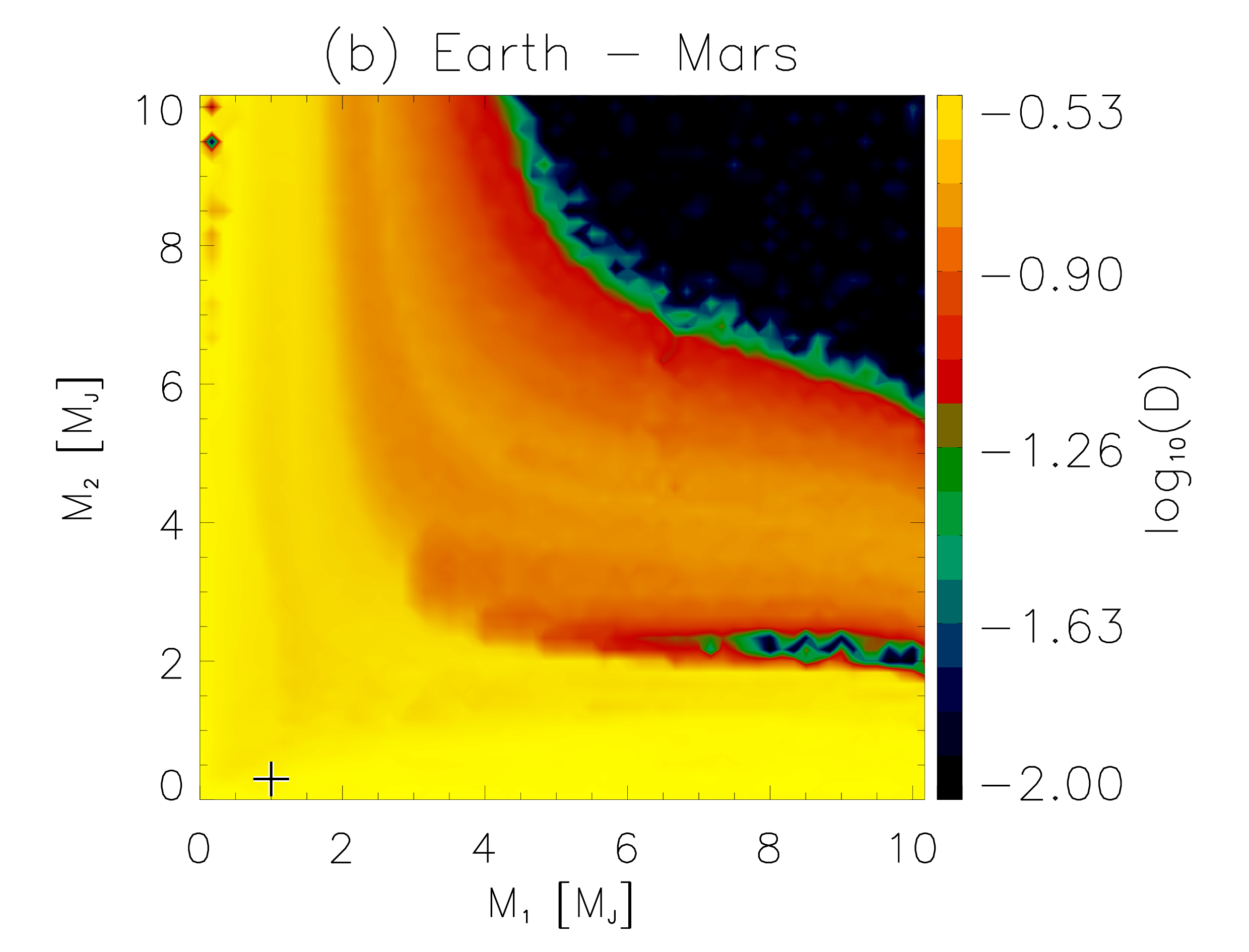}
\end{minipage}
\begin{minipage}{18.5pc}
\includegraphics[width=18.5pc]{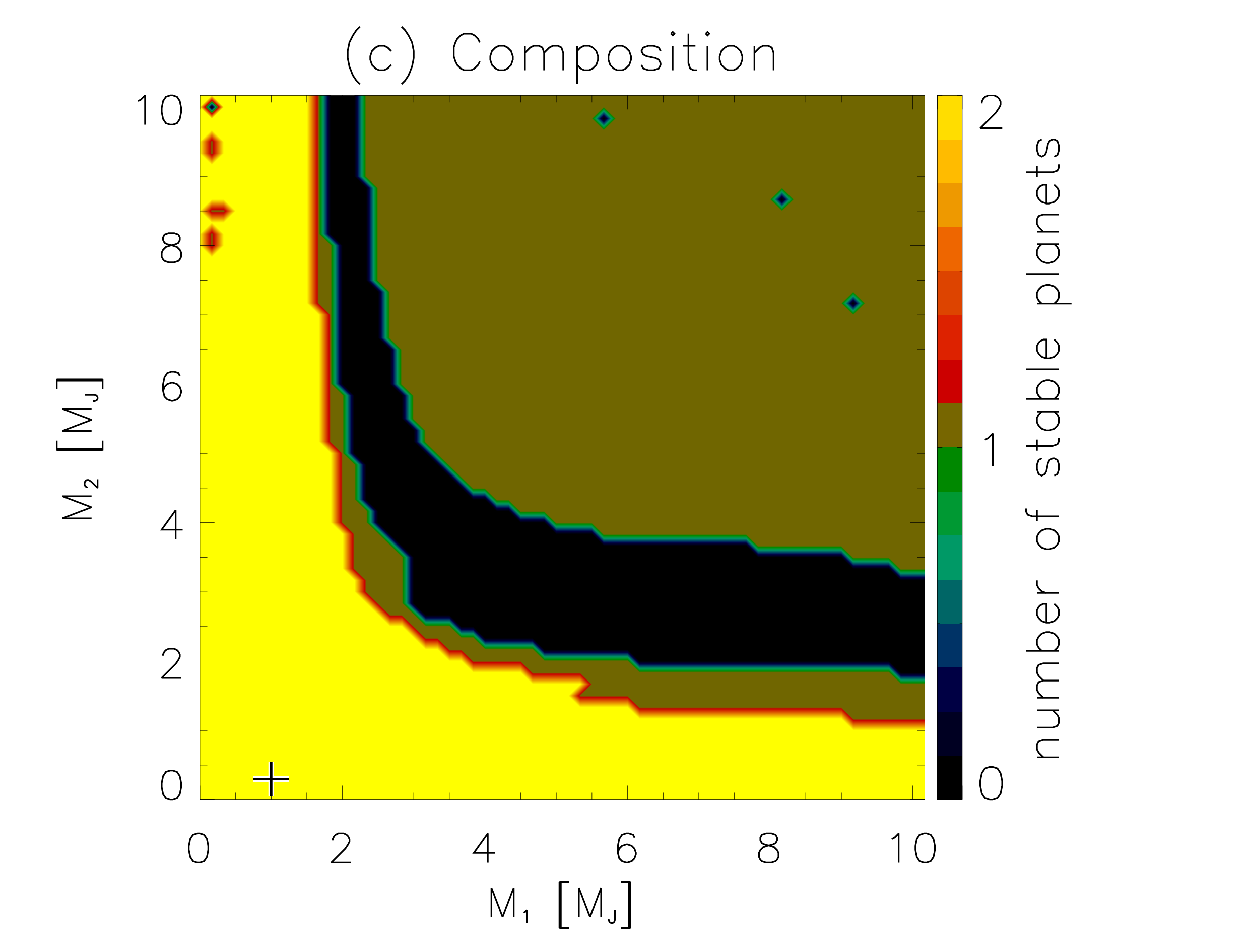}
\end{minipage}
\begin{minipage}{18.5pc}
\includegraphics[width=18.5pc]{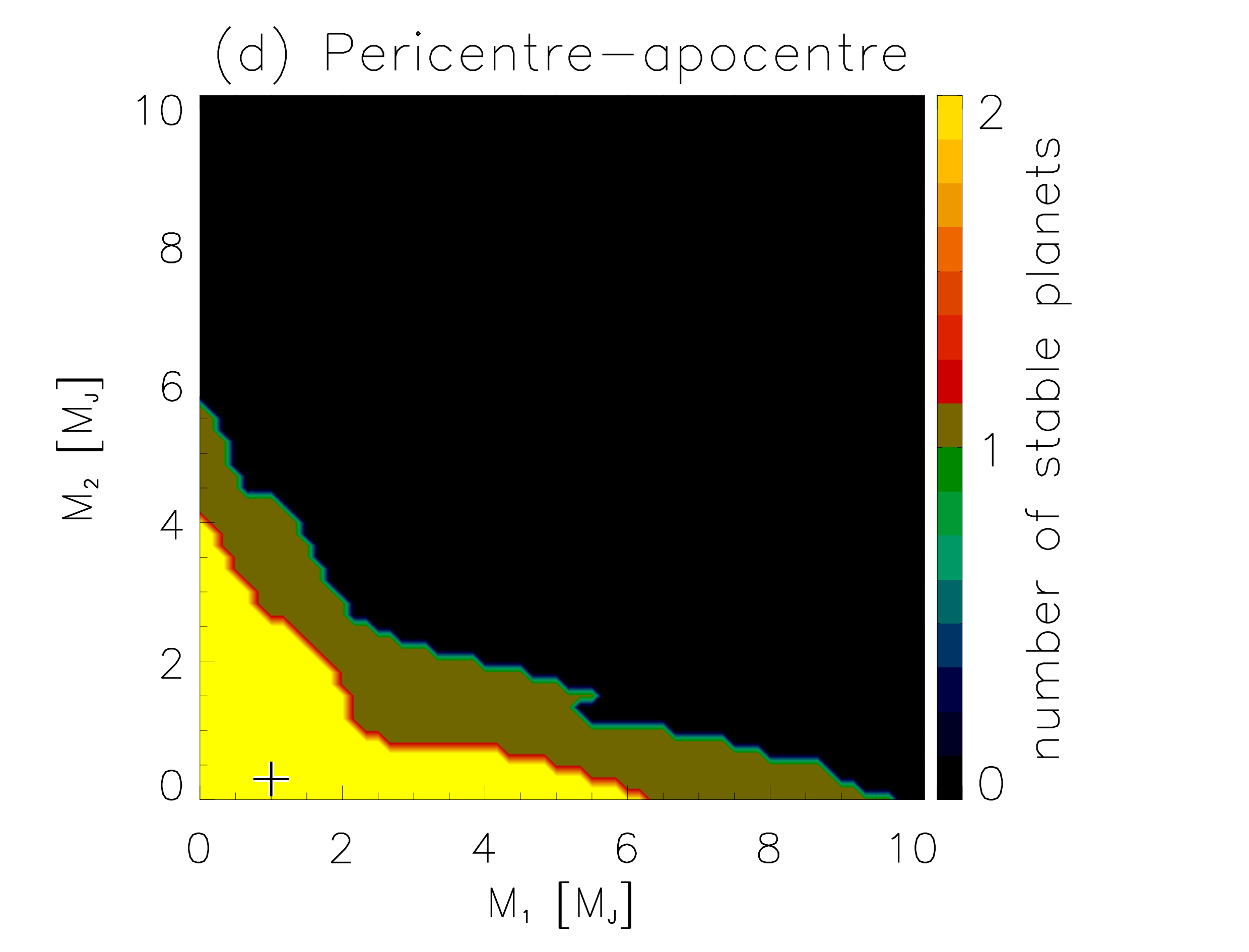}
\end{minipage}
\caption{\label{tav}($a$) and ($b$): Stability maps from close encounters. Yellow and black colours indicate the most stable and the most unstable regions, respectively. ($c$): Composition of stability maps seen in ($a$) and ($b$). ($d$): Stability map made from the difference between the minimum pericentre of the outer planet and the maximum apocentre of the inner planet. Colours indicate the number of stable planet pairs: yellow - two, brown - one, black - zero. The '+' mark indicates our Solar System.}

\end{figure}

Stability is not only determined by maximum eccentricities, but close encounters were also examined. To characterize the vicinity of neighbouring rocky planets the difference between the locations of these planets was calculated at every integration step. From these values the smallest was chosen as a measure of the encounter ($D$). To get comparable results for the two planet pairs ({\it Venus -- Earth} and {\it Earth -- Mars}) these values were normalized by the mean of the original semimajor axes:

\begin {equation}
D = \frac { min\{encounters\} } { \frac {a_1 + a_2} {2} } , \nonumber
\end {equation}
\\
where $a_1$ and $a_2$ are the semimajor axes of the closer and the farther planet from the Sun. The value of $D$ for different mass-pairs of giant planets can be seen in Fig. \ref{tav} ($a$) and ($b$). Yellow and black colours indicate the least close and the closest encounters, respectively. The {\it yellow -- orange} boundaries can be considered as the margin of instability. In the case of Venus and Earth encounters are only seen for mass values $M_1 \ge$ 6 $M_J$ and $M_2 \approx$ 2 $M_J$. But at Earth and Mars at the top right corner (at great masses of the giant planets) a huge black region can be seen caused by the instability of Mars mentioned above. In this area in several cases the semimajor axis of Mars is about 1 -- 2 $AU$ and its eccentricity almost reaches the value of 1.

From these stability maps a composite figure was made (see Fig. \ref{tav} ($c$)). The planet pair is considered unstable if the ratio of the value of the encounter and the mean of the original semimajor axes is less than 0.2. In Fig. \ref{tav} ($c$) the number of stable planet pairs can be seen. Yellow and black colours indicate the case when both and neither planet pairs are stable, respectively. To characterize the vicinity of the orbits of neighbouring rocky planets the difference between the outer planet's minimum pericentre distance and the inner planet's maximum apocentre distance was calculated as well. The orbits of the planet pairs are considered unstable if the ratio of the difference mentioned above and the mean of original semimajor axes is less than 0.01 (i.e. the planets come very close to each-other). In Fig. \ref{tav} ($d$) the number of planet pairs with stable orbits can be seen. This is the worst case that can be imagined, when the outer and the inner planets are located in their pericentre and apocentre, respectively, and these points and the Sun lie in a line. This may occur but the most probable case may be somewhere between the cases depicted in panel ($c$) and ($d$). In panel ($d$) yellow colour indicates that the orbits are surely stable while in panel ($c$) it indicates that they may be stable. Even in the worst case the stable (yellow) region is considerably large, hence Jupiter could be four times and Saturn could be three times more massive while the orbits of the inner planets stay stable.

\begin{figure}[h]
\centering

\begin{minipage}{38pc}
\includegraphics[width=38pc]{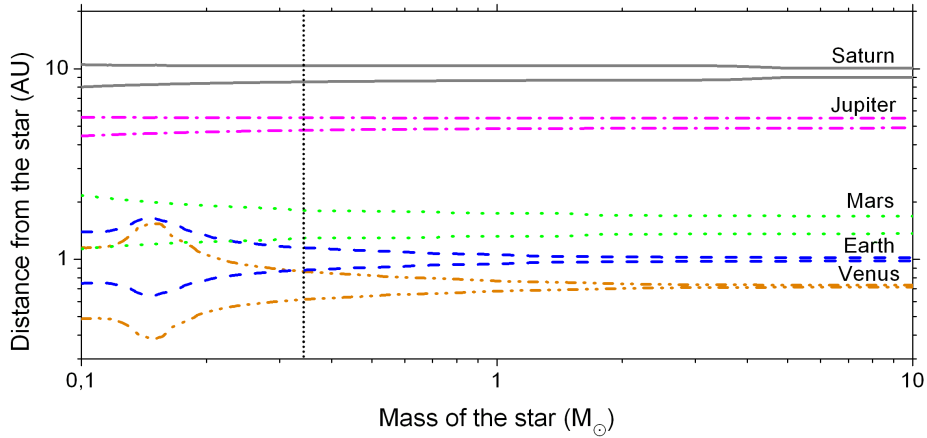}
\end{minipage}
\caption{\label{HZ1} The minimum pericentres and the maximum apocentres of the planets in function of the mass of the star. Different colours indicate different planets. The vertical spotted line at 0.33 $M_\odot$ is a limit for the stability: for less mass values the orbits of the planets are considered unstable and for greater values the orbits are considered stable.}

\end{figure}

The other problem that was investigated is also related to the stability of the inner planets but in this case the masses of the gas giants were constants while the mass of the Sun was altered between 0.1 $M_\odot$ and 10 $M_\odot$. The minimum pericentre and the maximum apocentre distances of the planets are plotted in Fig. \ref{HZ1} in function of the mass of the Sun. At low star masses the orbits of the inner planets can overlap each-other, hence two or even all three planets can encounter. From 0.33 $M_\odot$ (vertical spotted line on the figure) the orbits of the planets are considered stable regarding to the above-mentioned criterion (i.e. the ratio of the value of the encounter and the mean of the original semimajor axes of Venus and Earth is greater than 0.01).

In order to better understand the feature of habitability the two kinds of habitable zones were calculated. The results of these investigations are to be thoroughly discussed in the next chapter.

\subsection{Habitable zones}
\begin{figure}[h]
\centering

\begin{minipage}{18pc}
\includegraphics[width=18pc]{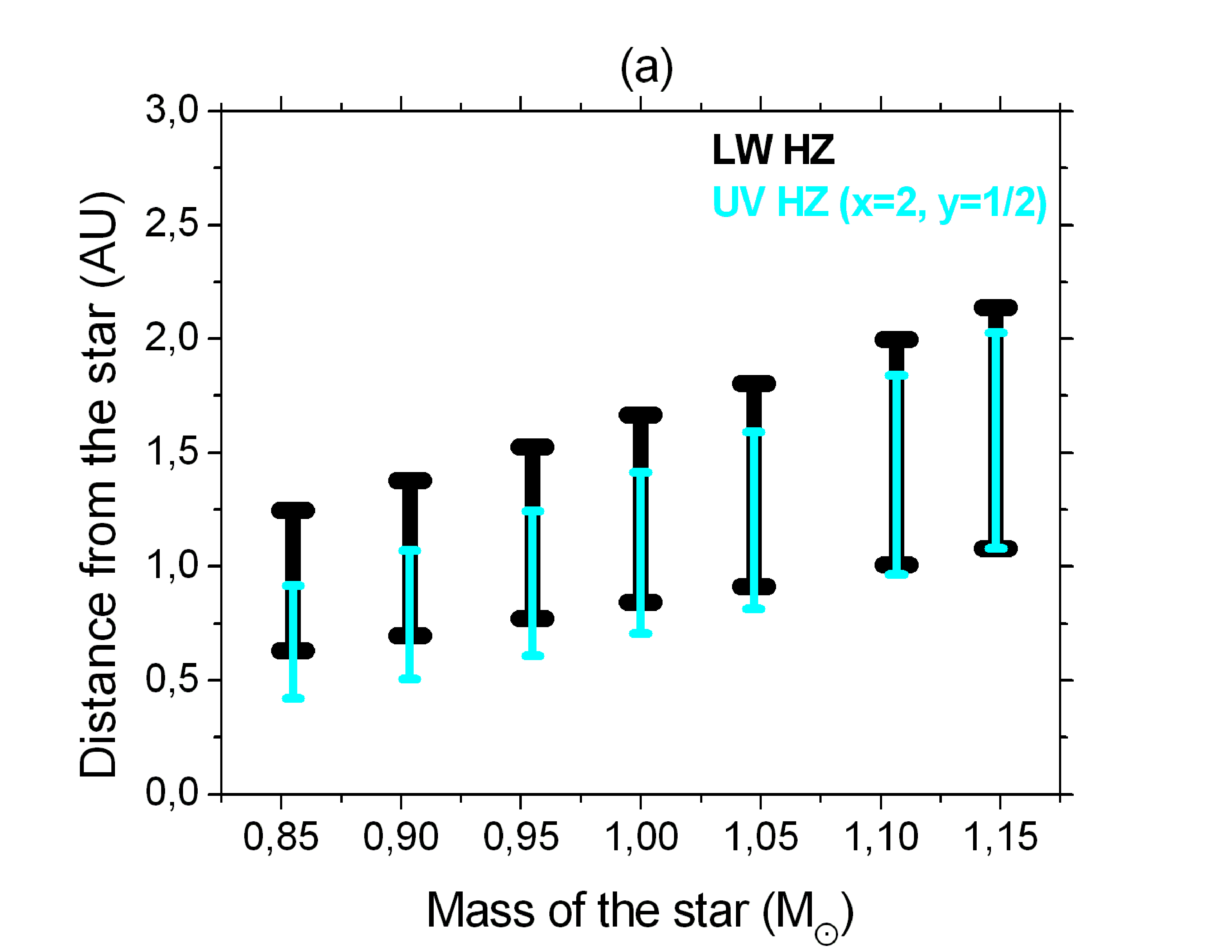}
\end{minipage}
\begin{minipage}{18pc}
\includegraphics[width=18pc]{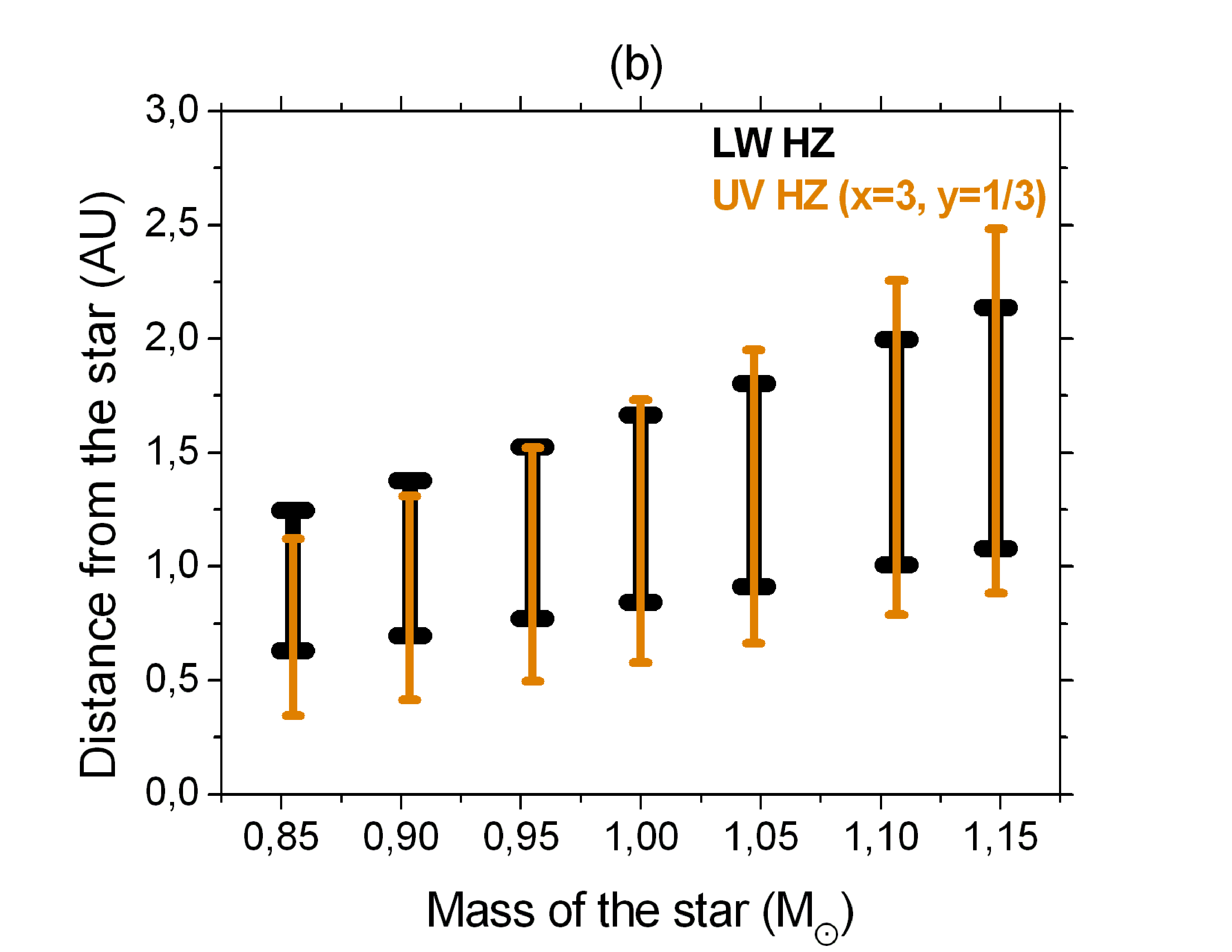}
\end{minipage}
\caption{\label{HZ2} The extension of the habitable zones in function of the mass of the star. ($a$): Black and light blue lines indicate the conventional and the UV habitable zone ($x$ = 2, $y$ = 1/2), respectively. ($b$): Black and orange lines indicate the conventional and the UV ($x$ = 3, $y$ = 1/3) habitable  zones, respectively.}

\end{figure}

The boundaries of habitable zones were calculated for seven values of mass in the 0.8 -- 1.2 $M_\odot$ interval using the definition of the LW HZ given by Kasting et al. (1993) and equations (1) and (2) for the UV HZ. Two $x$,$y$ sets were used for calculating the boundaries of the UV HZ. The first was $x$ = 2 and $y$ = 1/2 which was used by Buccino et al. (2006) as well. These are estimated values since different compounds and organisms have different level of tolerance (Cockell, 1998). As the values of these constants are difficult to estimate the extension of the zone was also investigated in the case of $x$ = 3 and $y$ = 1/3. The extension of both habitable zones for the seven masses of the star can be seen in panel ($a$) and ($b$) of Fig. \ref{HZ2}. In the first case ($x$ = 2 and $y$ = 1/2) the UV HZ has a similar extension to the conventional HZ and they overlap each-other, but the UV HZ is located slightly closer to the Sun in the 0.8 -- 1.2 $M_\odot$ interval. In the other case the UV HZ covers a wider range than the conventional HZ, especially the outer boundary moved farther. For higher masses than 0.95 $M_\odot$ the LW HZ is fully located inside the UV HZ. It can be seen that the Earth is located in both habitable zones even for lower and higher masses of the Sun. 

\begin{figure}[h]
\centering

\begin{minipage}{38pc}
\includegraphics[width=38pc]{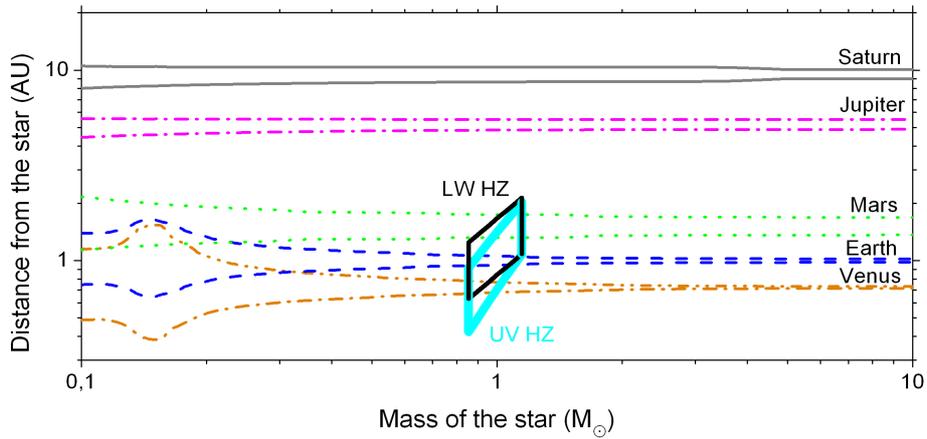}
\end{minipage}
\caption{\label{HZ3}  The minimum pericentres and the maximum apocentres of the planets like in Fig. \ref{HZ1}. Between  0.8 and 1.2 $M_\odot$ the extension of the habitable zones can be seen: black and light blue lines indicate the conventional and the UV habitable zone ($x$ = 2, $y$ = 1/2), respectively.}

\end{figure}

For detailed results see Fig. \ref{HZ3}, where black and light blue colours indicate again the conventional and the UV ($x = 2$ and $y = 1/2$) habitable zones, respectively. At lower masses of the Sun only Venus, but for higher mass values Earth and also Mars are located in both HZ.

\section{Conclusions}
In the celestial mechanics discussion the stability of the orbits of the rocky planets was investigated for different masses of the two gas giants. The inner planetary system remained stable for several mass pairs, hence the fine-tune problem does not prevail. Yet, long-term investigation would be worthwhile because Batygin and Laughlin (2008) found that Mars can escape from the Solar System in 800 -- 1200 million years.

In the other case when the mass of the Sun was altered, the orbits of the planets were stable for values greater than 0.33 $M_\odot$. Between 0.8 and 1.2 $AU$ the extensions of the two habitable zones were calculated and in almost all cases the Earth was located in both habitable zones. This result also suggests that the Solar System is not considered to be unique.

\ack
We thank Dr. B\'alint \'Erdi for useful comments and discussion.

\section*{References}
\begin{thereferences}
\item Batygin K and Laughlin G 2008 On the dynamical stability of the Solar System {\it ApJ} {\bf 683} 1207-16
\item Buccino A, Lemarchand G and Mauas P 2006 Ultraviolet radiation constraints around the circumstellar habitable zones {\it Icarus} {\bf 183} 491-503
\item Cockell C 1998 Biological effects of high ultraviolet radiation on early Earth - a theoretical evaluation {\it J. Theor. Biol.} {\bf 193} 717-29
\item Coohill T P 1991 Action spectra again? {\it Photochem. Photobiol.} {\bf 54} 859-70
\item Csizmadia S 2010 Notes on exoplanets {\it J. Phys.: Conf. Series} in this volume
\item Gough D O 1981 Solar interior structure and luminosity variations {\it Solar Physics} {\bf 74} 21-34
\item Horneck G 1995 Quantification of the biological effectiveness of environmental UV radiation {\it J. Photochem. Photobiol.} B: Biol. {\bf 31} 43-9
\item Kasting J, Whitmire D and Reynolds R 1993 Habitable zones around main sequence stars {\it Icarus} {\bf 101} 108-28
\item Modos K, Gaspar S, Kirsch P, Gay M and Ronto G 1999 Construction of spectral sensitivity function using polychromatic UV sources {\it J. Photochem. Photobiol.} B: Biol. {\bf 49} 171-6
\item Zaninetti L 2008 Semi-analytical formulas for the Hertzsprung-Russel Diagram {\it Serbian Astronomical Journal} {\bf 177} 73-85

\end{thereferences}

\end {document}